\documentclass[%prl,
 aps,
 ,twocolumn
]{revtex4}
\usepackage{graphicx}
\usepackage{amsmath}
\usepackage{amssymb}
\usepackage{mathrsfs}
\usepackage{color}
\usepackage{tikz}
\usepackage{physics}

\begin{document}

\title{Pump Depletion in Parametric Amplification}
\author{Wanli Xing}
\affiliation{Centre for Quantum Computation and Communication Technology, School of Mathematics and Physics, University of Queensland, St Lucia, Queensland 4072, Australia}
\author{T.C.Ralph}
\affiliation{Centre for Quantum Computation and Communication Technology, School of Mathematics and Physics, University of Queensland, St Lucia, Queensland 4072, Australia}
\begin{abstract}
{We derive analytic solutions for Heisenberg evolution under the trilinear parametric Hamiltonian which are correct to second order in the interaction strength but are valid for all pump amplitudes. The solutions allow pump depletion effects to be incorporated in the description of parametric amplification in experimentally relevant scenarios and the resulting new phenomena to be rigorously described.}
\end{abstract}
\maketitle

{\it Introduction}:
Optical parametric amplification \cite{Walls_2008} is the work-horse of quantum optics, being a source for single photons in the weak amplification regime \cite{Hong_1986, JIN14} and a source of squeezed states in the strong amplification regime \cite{Wu_1987,EBE13}.  A simple description of this interaction for the non-degenerate (two-mode squeezing) case is given by the unitary \cite{Bachor_Ralph_2019}
\begin{equation}
    U = \exp{-i\chi (a^\dag b^\dag \alpha + ab \alpha^*)},
    \label{uquad}
\end{equation}
where $a$ and $b$ are annihilation operators describing the squeezed modes, $\chi$ is the interaction strength and $\alpha$ is the amplitude of the coherent pump field. Applying this unitary to the vacuum produces the well known two-mode squeezed state
\begin{equation}
  \ket{\lambda}_{ab} =  U|0\rangle = \sqrt{1-\lambda^2} \sum \lambda^n \ket{n}_{a} \ket{n}_{b}
\end{equation}
where $\lambda = \tanh{\chi \alpha}$. Alternatively the Heisenberg evolution of the annihilation operators is given by
\begin{eqnarray}
a_0 &=  U^{\dagger} a U = a \;\cosh{\chi \alpha} - i b^{\dagger}\; \sinh{\chi \alpha} \nonumber \\
b_0 &=  U^{\dagger} b U = b \; \cosh{\chi \alpha} - i a^{\dagger}\; \sinh{\chi \alpha}.
\label{Hquad}
\end{eqnarray}
Being quadratic in the operators the squeezing unitary is Gaussian, i.e. mapping Gaussian states to Gaussian states and so the first and second moments of the Heisenberg operators and their Hermitian conjugates are sufficient to completely characterise squeezed states \cite{Weedbrook_2012}.\\

Sophisticated models based on this interaction can be built that successfully describe a large range of devices and protocols in quantum optics \cite{Baumgartner_1979,Dodonov_2002}, quantum communication \cite{Ou_1992}, quantum computing \cite{MEN06} and quantum metrology \cite{Clerk_2010, hudelist_kong_liu_jing_ou_zhang_2014}. Yet at a fundamental level this interaction is {\it unphysical} as it is not energy conserving. This is because the pump laser is treated as a reservoir that is unaffected, i.e. undepleted, by the interaction. Under typical experimental conditions this is a good approximation as the efficiency of the interaction is very low, however efficiencies are improving all the time and experiments are moving into the regime where depletion effects cannot be neglected \cite{ALL14, FLO20}. Whilst full numerical solutions have been known for many years \cite{Walls_1970, Drobny_1992} and have been used in theoretical studies \cite{NAT10, BIR20} in conjunction with short time perturbative approaches, these rapidly become intractable when treating realistic systems where the pump power is large.\\

In this work we derive Heisenberg equations of motion which include the lowest order non-trivial corrections to the standard equations due to pump depletion in a consistent manner that allows for large pump powers. Although non-linear in the mode operators, our equations are straightforward to work with and allow an exploration of the novel physics that arises and description of the most accessible experimental signatures of pump depletion.\\

{\it Heisenberg Evolution by the Trilinear Hamiltonian}:
The exact unitary describing the parametric amplification process is given by
\begin{equation}
    U = \exp{-i\chi(a^\dag b^\dag c + abc^\dag)}.
\end{equation}
Notice that the approximation that leads us back to quadratic form in Eq.\ref{uquad} is the replacement $c \to \langle c \rangle = \alpha$. We want to know the full form of operators $a, b, c$ in the Heisenberg picture, however we no longer obtain the simple closed form linear equations of Eq.\ref{Hquad}. Nevertheless 
they can be evaluated to any desired order using the Baker-Campbell-Hausdorff formula \cite{sakurai_napolitano_2017},
\begin{align}
    a_o & = e^{G} a e^{-G} \nonumber \\
    & = a + [G,a] + \tfrac{1}{2!} [G,[G,a]] + \tfrac{1}{3!}[G,[G,[G,a]]] + \dots
\end{align}
%\section{Simplifications}
The Heisenberg operators get very complicated as we include terms of higher and higher orders. The evolved operators up to order $\chi^8$ are given explicitly in Appendix A. However, we find that a brute force expansion in orders of $\chi$ is not the most tractable approach in situations of experimental interest.
This is because, for typical experimental parameters, not all terms with the same power of $\chi$ contribute equally when we calculate expectation values. To see this we can write the pump operator as $\hat c = \alpha + {\hat \delta c}$, where the expectation value of $\hat c$, $\langle \hat c \rangle = \alpha$, is the coherent amplitude of the pump, which we assume to be real here; $\hat{\delta c}$ is an operator representing the noise/quantum part of the pump; and $\langle{\hat{\delta c}}\rangle = 0$ by definition. If we carry out this expansion, then, for example, the first few terms of $a_o$ become:
\begin{align}
    & a_o = a - i\chi \alpha b^\dag - i\chi b^\dag \delta c \nonumber \\ 
    & + \tfrac{\chi^2}{2!} \qty(-ab^\dag b + a\alpha^2 + \alpha a\delta c^\dag + \alpha a \delta c + a\delta c^\dag \delta c) + \dots 
\end{align}
$\alpha$ can be much larger than 1, whereas $\chi$ is a small number much less than 1. We assume $\alpha\chi$ is of order unity. In this case we see that terms like $-\tfrac{\chi^2}{2!}ab^\dag b$, which is of order $\chi^2$, will contribute much less than terms like $\tfrac{\chi^2}{2!} \alpha^2 a$, which is of order $(\alpha\chi)^2 \sim 1$. The result is that we need to consider both $\alpha$ and $\chi$ when doing the expansion, and the size of a term is determined by the difference in powers between $\alpha$ and $\chi$. In the end we wish to derive consistent Heisenberg operator equations which can be used to evaluate first, second and third order expectation values that are accurate to second order in $\chi$ and to all orders in $\chi \alpha$.\\

Hence we perform the $\hat c = \alpha + \hat{\delta c}$ expansion, and only keep terms of the form $\alpha^n\chi^n$ and $\alpha^{n-1}\chi^n$, and ignore any other terms (there are also no terms where the power of $\alpha$ is higher than the power of $\chi$). Then the operator for the signal looks like
\begin{align}
    a_o & = a - i\alpha\chi  b^\dag - i\chi b^\dag \delta c + \tfrac{\chi^2}{2!} \qty( \alpha^2 a +  \alpha a \delta c^\dag + \alpha a \delta c)\nonumber \\
    & + \tfrac{i\chi^3}{3!}\qty(-    \alpha ^ 3  b^{\dag} -     \alpha ^ 2  b^{\dag}  \delta c^{\dag} - 2      \alpha ^ 2  b^{\dag}\delta c ) \nonumber \\
    & + \tfrac{\chi^4}{4!}\qty( \alpha ^ 4  a + 2   \alpha ^ 3  a  \delta c^{\dag} + 2   \alpha ^ 3  a\delta c ) \nonumber \\
    & + \tfrac{i\chi^5}{5!}\qty(-    \alpha ^ 5  b^{\dag} - 2      \alpha ^ 4  b^{\dag}  \delta c^{\dag} - 3      \alpha ^ 4  b^{\dag}\delta c ) \nonumber \\
    & + \tfrac{\chi^6}{6!}\qty( \alpha ^ 6  a + 3   \alpha ^ 5  a  \delta c^{\dag} + 3   \alpha ^ 5  a\delta c ) \nonumber \\
    & + \tfrac{i\chi^7}{7!}\qty(-    \alpha ^ 7  b^{\dag} - 3      \alpha ^ 6  b^{\dag}  \delta c^{\dag} - 4      \alpha ^ 6  b^{\dag}\delta c ) \nonumber \\
    &+ \tfrac{\chi^8}{8!}\qty( \alpha ^ 8  a + 4   \alpha ^ 7  a  \delta c^{\dag} + 4   \alpha ^ 7  a\delta c ) + \dots  %\\ 
    \end{align}
Collecting terms with the same operators, we obtain several expansions in $\alpha\chi$:
\begin{align}
    a_o & = a (1 + \tfrac{\alpha^2\chi^2}{2!} + \tfrac{\alpha^4\chi^4}{4!} + \tfrac{\alpha^6\chi^6}{6!} + \tfrac{\alpha^8\chi^8}{8!} + \dots) \nonumber \\
    & + a(\delta c + \delta c^\dag) (\tfrac{\alpha\chi^2}{2!} + \tfrac{2\alpha^3\chi^4}{4!} + \tfrac{3\alpha^5\chi^6}{6!} + \tfrac{4\alpha^7\chi^8}{8!} + \dots) \nonumber \\
    & - ib^\dag (\alpha\chi + \tfrac{\alpha^3\chi^3}{3!} + \tfrac{\alpha^5\chi^5}{5!} + \tfrac{\alpha^7\chi^7}{7!} + \dots) \nonumber \\
    & - ib^\dag \delta c (\chi + \tfrac{2\alpha^2\chi^3}{3!} + \tfrac{3\alpha^4\chi^5}{5!} + \tfrac{4\alpha^6\chi^7}{7!} + \dots)  \nonumber \\
    & -ib^\dag \delta c^\dag (\tfrac{\alpha^2\chi^3}{3!} + \tfrac{2\alpha^4\chi^5}{5!} + \tfrac{3\alpha^6\chi^7}{7!} + \dots).
\end{align}
We see a clear pattern from each of the expansions. Assuming the pattern persists (checked to order $\chi^{15}$), we can write the terms in each bracket as an infinite sum, which are found to have closed form expressions. The coefficients of each operator are:
%\begin{equation}
%\begin{alignedat}{7}
\begin{align}
    % alternative aligning: & a: && sum.....
    a: \qquad
    % \ 1 + \tfrac{\alpha^2\chi^2}{2!} + \tfrac{\alpha^4\chi^4}{4!} + \tfrac{\alpha^6\chi^6}{6!} + \dots 
    &  \sum_{n=0}^\infty \tfrac{\alpha^{2n}\chi^{2n}}{(2n)!} = \cosh\alpha\chi, \nonumber
     \\
    a(\delta c + \delta c^\dag):
    %\ \tfrac{\alpha\chi^2}{2!} + \tfrac{2\alpha^3\chi^4}{4!} + \tfrac{3\alpha^5\chi^6}{6!} +  \dots 
    &  \sum_{n=1}^\infty \tfrac{\alpha^{2n-1}\chi^{2n}n}{(2n)!}  = \tfrac{\chi}{2}\sinh\alpha\chi, \nonumber \\
     -ib^\dag: \qquad
    %\ -i\qty(\alpha\chi + \tfrac{\alpha^3\chi^3}{3!} + \tfrac{\alpha^5\chi^5}{5!} +  \dots) 
    &   \sum_{n=0}^\infty \tfrac{\alpha^{2n+1}\chi^{2n+1}}{(2n+1)!} =  \sinh \alpha\chi, \nonumber \\
     -b^\dag \delta c: \qquad
    %\ -i\qty(\chi + \tfrac{2\alpha^2\chi^3}{3!} + \tfrac{3\alpha^4\chi^5}{5!} +  \dots) 
    &   \sum_{n=0}^\infty \tfrac{\alpha^{2n}\chi^{2n+1}(n+1)}{(2n+1)!} \nonumber \\
     & \quad = \tfrac{i\chi}{2}\cosh\alpha\chi + \tfrac{i}{2\alpha}\sinh\alpha\chi, \nonumber \\
     -b^\dag \delta c^\dag: \qquad
    % \ -i\qty(\tfrac{\alpha^2\chi^3}{3!} + \tfrac{2\alpha^4\chi^5}{5!} + \tfrac{3\alpha^6\chi^7}{7!} + \dots) 
    &  \sum_{n=1}^\infty \tfrac{\alpha^{2n}\chi^{2n+1}n}{(2n+1)!} \nonumber  \\
     & \quad = \tfrac{i\chi}{2}\cosh\alpha \chi - \tfrac{i}{2\alpha}\sinh\alpha\chi.
%    \end{alignedat}
%\end{equation}
\label{eqn:aoExpansion}
\end{align}
These terms give us expressions that are valid up to order $\chi$. Now, we are interested in the second order moments which will be to order $\chi^2$, hence to be safe we should expand to order $\chi^2$ in a similar way to what we have done when expanding to order $\chi$. This full expansion is performed in Appendix. B %\ref{appx:orderChi2Ops}. 
However, the vast majority of the second order terms derived in the appendix do not contribute to the expectation values at $\mathcal O(\chi^2)$. 
%Specifically, it turns out the only second order term in $a_o$ that contributes to $\expval*{a_o^\dag a_o}$ is $\chi^2 A_{b^\dag}b^\dag$; the only second order terms that contribute to $\expval*{a_ob_o}$ are $\chi^2(A_{b^\dag}b^\dag + A_aa)$. Similarly, the only second order or above term in $c_o$ that contributes to $\expval*{c_o^\dag c_o}$ is $\chi^3 C_\alpha $, and the only second order term that contribute to $\expval*{\delta c_o\delta c_o}$ is $\chi^2C_{\delta c^\dag}\delta c^\dag$. All other terms either annihilate $\bra0$ or $\ket0$ to give 0 contributions, or they only contribute to $\mathcal O(\chi^3)$ terms. 
We find that for the purpose of calculating expectation values, we may take
\begin{align}
    a_o & = a \qty[\cosh\chi^\prime + (\delta c + \delta c^\dag) \tfrac{\chi}{2}\sinh\chi^\prime] \nonumber \\
    & - ib^\dag \big[\sinh \chi^\prime + \tfrac{\chi}{2}\cosh\chi^\prime(\delta c + \delta c^\dag) \nonumber \\
    & \qquad + \tfrac{\chi}{2\chi^\prime} \sinh\chi^\prime(\delta c - \delta c^\dag) \big] \nonumber \\
    & \qquad + \chi^2(i A b^\dag + B a), 
    \label{signal}
   \end{align}
   \begin{align}
    b_o & = b \qty[\cosh\chi^\prime + (\delta c + \delta c^\dag) \tfrac{\chi}{2}\sinh\chi^\prime] \nonumber \\
    & - ia^\dag \big[\sinh \chi^\prime + \tfrac{\chi}{2}\cosh\chi^\prime(\delta c + \delta c^\dag) \nonumber \\
    & \qquad + \tfrac{\chi}{2\chi^\prime} \sinh\chi^\prime(\delta c - \delta c^\dag) \big] \nonumber \\
    & \qquad + \chi^2( B b + i A  a^\dag), 
    \label{idler}
   \end{align}
      \begin{align}
    c_o &=\alpha_o + \delta c_o \nonumber \\
    & = \alpha - \tfrac{\chi}{2\chi^\prime}\sinh^2\chi^\prime + \chi^3 C \nonumber\\
    & \quad + \delta c  - (a^\dag a + b^\dag b) \tfrac{\chi}{2\chi^\prime}\sinh^2\chi^\prime \nonumber\\
    & \quad - ia^\dag b^\dag \tfrac{\chi}{2}(1 - \tfrac{1}{\chi^\prime}\sinh\chi^\prime \cosh\chi^\prime) \nonumber\\
    & \quad - iab \tfrac{\chi}{2} (1 + \tfrac{1}{\chi^\prime}\sinh\chi^\prime\cosh\chi^\prime) \nonumber \\
    & \quad + \chi^2 D \delta c^\dag.
    \label{pump}
\end{align}
where
\begin{align*}
    A & =  \tfrac{-5\chi^\prime\cosh\chi^\prime+2\sinh\chi^\prime-\chi^{\prime2}\sinh\chi^\prime+\sinh3\chi^\prime}{8\chi^{\prime2}},\\
    B & = -\tfrac{-\cosh\chi^\prime-\chi^{\prime2}\cosh\chi^\prime+\cosh3\chi^\prime-3\chi^\prime\sinh\chi^\prime}{8\chi^{\prime2}}, \\
    C & =  \tfrac{-3-4\chi^{\prime2}+(2-4\chi^{\prime2})\cosh2\chi^\prime + \cosh4\chi^\prime-2\chi^\prime\sinh2\chi^\prime}{32\chi^{\prime3}} \\
  D  & =  - \tfrac{1-\cosh2\chi^\prime+\chi^\prime\sinh2\chi^\prime}{4\chi^{\prime2}}.
\end{align*}
and $\chi' = \alpha \chi$. These are `effective operators' in the sense that they give the correct results for all normally ordered second order moments, i.e. $\expval*{a_o^\dag a_o}, \expval*{a_ob_o}, \expval*{\delta c_o^\dag \delta c_o}, \expval*{\delta c_o\delta c_o}$ and $\alpha_o^2$, and therefore all variances calculated from these operators are correct. Terms that arise in any calculation that are not normally ordered must first be reordered using the standard Boson commutator relations, e.g. $[a_o, a_o^\dag] = 1$, before proceeding with the calculation. \\

\indent Eqs \ref{signal}, \ref{idler} and \ref{pump} are the main results of this paper. They provide a tractable and physically intuitive way to investigate the lowest order corrections to the behaviour of two-mode squeezing when pump depletion becomes significant.\\

{\it Expectation values}:
We are now in a position to investigate the physics of the pump-depleted squeezer. Let us first consider the photon number in the pump and the squeezed modes. Assuming the pump is initially in a coherent state and the squeezed modes are initially in vacuum states, the photon number in the pump after the interaction is given by
\begin{align}
    \expval*{c^\dag_oc_o} & = \alpha^2_o + \expval*{\delta c^\dag_o\delta c_o} \nonumber \\
    &= \alpha^2 - \sinh^2 \chi' + (\tfrac{\chi}{2\chi^\prime})^2\sinh^4\chi^\prime + 2 \chi^2 \chi^\prime C \nonumber \\
    +  \tfrac{\chi^2}{4} & (1 - \tfrac{2}{\chi^\prime}\sinh\chi^\prime\cosh\chi^\prime + \tfrac{1}{\chi^{\prime2}}\sinh^2\chi^\prime\cosh^2\chi^\prime  ).
    \label{eqn:alphaAnddcDagdc}
\end{align}
As expected the pump is now depleted by the interaction with $ \expval*{c^\dag_oc_o}< \alpha^2$. In addition there is now a coherent contribution to the photon number, $\alpha_o^2$ and an incoherent contribution, $ \expval*{\delta c^\dag_o\delta c_o}$.
The photon numbers in the squeezed modes are given by
\begin{align}
    \expval*{a^\dag_oa_o} &= \expval*{b^\dag_ob_o} = \sinh^2\chi^\prime -2 \chi^2 \sinh\chi^\prime A \nonumber \\
    + \tfrac{\chi^2}{4}& (\cosh^2\chi^\prime - \tfrac{2}{\chi^\prime}\sinh\chi^\prime\cosh\chi^\prime + \tfrac{1}{\chi^{\prime2}}\sinh^2\chi^\prime).
    \label{eqn:aDaga}
\end{align}
The photon number in the squeezed modes are also lower than that predicted by the undepleted pump model. It is straightforward to confirm that energy conservation now holds as
\begin{eqnarray}
      \expval*{c^\dag_oc_o} + {{1}\over{2}}( \expval*{a^\dag_oa_o} + \expval*{b^\dag_ob_o}) = \alpha^2 ,
\end{eqnarray}
where we have taken into account that the energy of the squeezed mode photons is half that of the pump photons.

The other non-zero expectation values up to third order can also be calculated and give
%Let the state be the vacuum, $\ket0_a\ket0_b\ket0_c$, we can now calculate the correlations between various operators, keep terms to order $\chi^2$, and use the subscript $o$ to label the time-dependent quantities, for example $a(t)\equiv a_o, \alpha(t) \equiv \alpha_o$, we have
%\begin{subequations}
\begin{align}
   %  \expval*{a_oa_o} & = 0, \\
    %
  %  \expval*{a^\dag_oa_o} &  = \sinh^2\chi^\prime + \tfrac{\chi^2}{4}\qty(\cosh^2\chi^\prime - \tfrac{2}% {\chi^\prime}\sinh\chi^\prime\cosh\chi^\prime + \tfrac{1}{\chi^{\prime2}}\sinh^2\chi^\prime),\nonumber \\
    %
    \expval*{a_o b_o}  & = -\tfrac{i}{2}\sinh2\chi^\prime \nonumber \\
    & +  \tfrac{i\chi^2}{16\chi^{\prime2}} \Big( -4\chi^\prime-6\chi^\prime\cosh2\chi^\prime\nonumber \\
    &\qquad + (1-4\chi^{\prime2})\sinh2\chi^\prime + 2\sinh4\chi^\prime \Big),\nonumber \\
    \expval*{\delta c_o\delta c_o}& = \tfrac{1}{32}\Big(-8\chi^2(1+\tfrac{1}{\chi^\prime}\sinh2\chi^\prime) \nonumber \\
    & \qquad + \tfrac{\chi^2}{\chi^{\prime2}}(8\cosh2\chi^\prime+\cosh4\chi^\prime-9)\Big), 
   \nonumber \\
    %
   %  \expval*{\delta c^\dag_o \delta c_o} & = \tfrac{\chi^2}{4} \qty(1 - \tfrac{2}{\chi^\prime}\sinh\chi^
   %\prime\cosh\chi^\prime + \tfrac{1}{\chi^{\prime2}}\sinh^2\chi^\prime\cosh^2\prime  ),\nonumber \\
    %
    \expval*{a_ob_o\delta c_o} & = \tfrac{i\chi}{2\chi^\prime}\sinh\chi^\prime \cosh^3\chi\prime-\tfrac{i\chi}{2}\cosh^2\chi^\prime , \nonumber\\
    \expval*{a_ob_o\delta c^\dag_o} & = \tfrac{i\chi}{2\chi^\prime}\sinh^3\chi^\prime\cosh\chi^\prime-\tfrac{i\chi}{2}\sinh^2\chi^\prime . 
\label{eqn:correlations}
\end{align}
%\end{subequations}
From these we can calculate other interesting observables such as the quadrature variances of the output pump beam. The amplitude variance is given by
\begin{align}\label{vxc}
V_{xc} & =\langle (\delta c_o + \delta c_o^{\dagger})^2 \rangle  \nonumber \\
& = 2  \expval*{\delta c^\dag_o \delta c_o} + \expval*{\delta c_o \delta c_o} + \expval*{\delta c_o^{\dagger} \delta c_o^{\dagger}} + 1 \nonumber \\  
 &=  1-\tfrac{\chi^2}{\chi^\prime}\sinh2\chi^\prime +\nonumber \\
 &\;\;\;\; \tfrac{\chi^2}{8\chi^{\prime2}}(-5+4\cosh2\chi^\prime+\cosh4\chi^\prime),
\end{align}
whilst the phase variance is given by% \textcolor{blue}{added minus sign on first line}
\begin{align}\label{vpc}
V_{pc} & =-\langle (\delta c_o - \delta c_o^{\dagger})^2 \rangle  \nonumber \\ 
&=2  \expval*{\delta c^\dag_o \delta c_o} - \expval*{\delta c_o \delta c_o} - \expval*{\delta c_o^{\dagger} \delta c_o^{\dagger}} + 1 \nonumber \\      & = 1 - \chi^2(\tfrac{\sinh^2\chi^{\prime}}{\chi^{\prime2}}-1). 
\end{align}
We notice that the output pump has $V_p < 1 < V_x$ indicating it has become phase squeezed through the interaction. Because $\expval*{a_oa_o}  =  \expval*{b_ob_o} = 0$ the quadrature variances of the signal and idler are isotropic and given by $V_{xj} = V_{pj} = 1 + 2  \expval*{j^\dag_oj_o}$ where $j =\{a,b\}$. Given the phase convention we have adopted the correlations between the signal and idler exist between orthogonal quadratures. Hence the difference and sum squeezing between the signal and idler are given by
\begin{align}
V_{xp}^{\pm} &={{\langle ( a_o +  a_o^{\dagger}\pm i(b_o^{\dagger} -  b_o))^2 \rangle}\over{2}} \nonumber \\
&= 2  \expval*{ a^\dag_o  a_o} \mp 2i \expval*{ a_o  b_o} %\pm i \expval*{ a_o^{\dagger}  b_o^{\dagger}} 
+ 1 \nonumber\\  & = \frac{1}{16}(\cosh\chi^\prime\mp\sinh\chi^\prime)\;\;\; \times \nonumber\\
    & \quad \bigg[- \tfrac{3\chi^2}{\chi^{\prime2}}\cosh3\chi^\prime +\nonumber \\
    & ( \tfrac{3\chi^2}{\chi^{\prime2}}\mp \tfrac{20\chi^2}{\chi^{\prime}}+8(2+\chi^2))\cosh\chi^\prime \nonumber \\
    & \quad  + 2(\pm\tfrac{5\chi^2}{\chi^{\prime2}}(1+\cosh2\chi^\prime)+\nonumber \\
    &\tfrac{2\chi^2}{\chi^\prime}\mp4(2+\chi^2))\sinh\chi^\prime\bigg].
\end{align}
We find $V_{xp}^{+} < 1 < V_{xp}^{-}$ indicating entanglement between the signal and idler beams as expected. 

In contrast to the undepleted case, neither the output pump or signal and idler are in minimum uncertainty states. That is, we find $V_{xc}V_{pc}>1$ and $V_{xp}^{+}V_{xp}^{-}>1$. Given the overall unitarity of the interaction this indicates either non-Gaussianity or entanglement or both is emerging between the pump and the signal and idler. Indeed it is both. There is no correlation between the pump or signal or idler in the second order moments as would be required if Gaussian entanglement was emerging. Instead we find correlation between the pump and signal and idler in the third order moments indicating non-Gaussian entanglement. In particular we can consider the quadrature correlations $V_{abc} = \langle X_{ax} X_{bp} \delta X_{cx} \rangle$ and show %\textcolor{blue}{(Added minus sign on second line)}
\begin{align}
& V_{abc} = \langle i(a_o+a_o^{\dagger})(b_o^{\dagger}-b_o)(\delta c_o+\delta c_o^{\dagger}) \rangle \nonumber \\
& = -2 i ( \expval*{a_o b_o \delta c_o} + \expval*{a_o b_o \delta c^\dag_o} ) \nonumber \\ 
& = \tfrac{\chi}{2}(\tfrac{1}{\chi^\prime}\cosh\chi^\prime\sinh\chi^\prime-1)(\cosh^2\chi^\prime+\sinh^2\chi^\prime),
%or:
%\\ & = \tfrac{\chi}{2\chi^\prime}(\cosh^3\chi^\prime\sinh\chi^\prime + \cosh\chi^\prime\sinh^3\chi^\prime)-\tfrac{\chi}{2}(\cosh^2\chi^\prime+\sinh^2\chi^\prime),
%
%&=-\tfrac{i\chi}{2}\cosh^2\chi^\prime + \tfrac{i\chi}{2\chi^\prime}\sinh\chi^\prime \cosh^3\chi^\prime -\tfrac{i\chi}{2}\sinh^2\chi^\prime + \tfrac{i\chi}{2\chi^\prime}\sinh^3\chi^\prime\cosh\chi^\prime. 
\end{align}
The fact that this moment is non-zero (whilst all related first order moments are zero) indicates a non-Gaussian quantum correlation, i.e. entanglement.\\
%and recall $\alpha_o = \alpha + \tfrac{\chi}{2\chi^\prime}\sinh^2\chi^\prime$. If we want to find correlations such as $\expval*{c_oc_o}$ (instead of only the noise part $\expval*{\delta c_o\delta c_o}$), we have
%\begin{subequations}
%\begin{align}
%    \expval*{c_oc_o} & = \expval*{(\alpha_o+\delta c_o)^2} = \alpha^2_o + \expval*{\delta c_o\delta c_o},
%\end{align}
%since $\expval*{\delta c_o} = 0$. Similarly,
%\begin{align}
%    \expval*{c^\dag_oc_o} & = \alpha^2_o + \expval*{\delta c^\dag_o\delta c_o}, \\
%    %
%    \expval*{a_ob_oc_o} & = \alpha_o\expval*{a_ob_o} + \expval*{a_ob_o\delta c_o}, \\
%    %
%    \expval*{a_ob_oc^\dag_o} & = \alpha_o\expval*{a_ob_o} + \expval*{a_ob_o\delta c^\dag_o}. 
%\end{align}
%\end{subequations}
\begin{figure}[ht!]
\includegraphics[width=8.0cm]{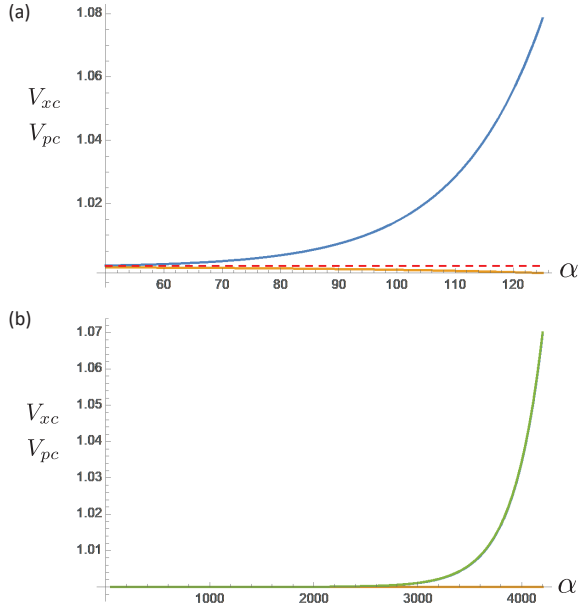}
\caption{\footnotesize Amplitude and phase variances of the output pump as a function of input pump amplitude $\alpha$: (a) the amplitude variance is shown in blue (upper) trace and the phase variance is the orange (lower) trace. The red-dashed line is the quantum noise limit. Here $\chi = 0.02$. A small amount of squeezing is seen for these parameters which are plotted using Eqs \ref{vxc} and \ref{vpc}; (b) the amplitude variance is shown in green (upper) trace and the phase variance is the orange (lower) trace. Here $\chi = 0.001$.  The stronger pump powers in this regime mean that now $V_{pc} = 1$ and $V_{xc}$ is given by Eqs \ref{vxcs}.} 
\label{GeneralCircuit}
\end{figure}
{\it Strong pump regime}: One parameter regime which is expected to be relevant for experimental tests of these effects is the strong pump regime. That is we take $\alpha$ sufficiently large that $\chi^\prime >> 1$, whilst still insisting $\chi$ is sufficiently small that our second order expansion remains valid. We note that although this regime is inaccessible to numerical approaches, it is easily explored with our analytical expressions. By neglecting the negative exponentials in our cosh and sinh terms and keeping only the largest of the positive exponentials we can significantly simplify our expectation values. The average photon numbers of the pump, signal and idler become:
\begin{align}
    \expval*{c^\dag_oc_o} & = \alpha^2  - {\tfrac{e^{2 \chi^\prime}}{4}} +\tfrac{\chi^2}{16 \chi^{\prime2}}\;\;e^{4 \chi^\prime} ,
    \label{eqn:alphaAnddcDagdcs}
\end{align}
and
\begin{align}
    \expval*{a^\dag_oa_o} &= \expval*{b^\dag_ob_o} =  {\tfrac{e^{2 \chi^\prime}}{4}} -\tfrac{\chi^2}{16 \chi^{\prime2}}\;\;e^{4 \chi^\prime} .
    \label{eqn:aDagas}
\end{align}
\begin{figure}[ht!]
\includegraphics[width=8.0cm]{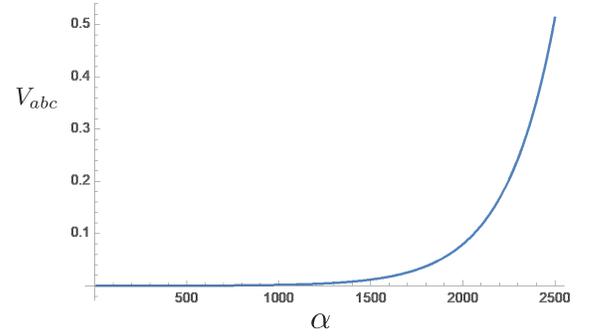}
\caption{\footnotesize Third order correlation $V_{abc}$ as a function of input pump amplitude $\alpha$. Here $\chi = 0.001$. Significant affects are seen at relatively low pump amplitude.} 
\label{GeneralCircuit}
\end{figure}
The pump amplitude quadrature variance becomes
\begin{align}\label{vxcs}
V_{xc} & = 1 +\tfrac{\chi^2}{16 \chi^{\prime2}}\;\;e^{4 \chi^\prime},
\end{align}
whilst the phase quadrature remains at the quantum noise level, $V_{pc} = 1$, given this approximation. The difference squeezing between the signal and idler is given by 
\begin{align}
V_{xp}^{+} &= e^{-2 \chi^\prime} + \tfrac{\chi^2}{16 \chi^{\prime2}}\;\;e^{2 \chi^\prime},
\end{align}
whilst the sum squeezing is given by 
\begin{align}
V_{xp}^{-} &= e^{2 \chi^\prime} - \tfrac{\chi^2}{4 \chi^{\prime2}}\;\;e^{4 \chi^\prime}.
\end{align}
Notice this leads to the uncertainty product $V_{xp}^{-} V_{xp}^{+} = 1 + {{\chi^2}\over{16 \chi^{\prime2}}}\;\;e^{{4 \chi^\prime}}$ %\textcolor{blue}{(full expansion: there's also a $-\tfrac{17 \chi^4e^{6\chi^\prime}}{256\chi^{\prime4}}$ term, although it is smaller than $\chi^2e^{4\chi^\prime}$ at short time)}, 
indicating the departure from a pure Gaussian entangled state.

Perhaps surprisingly the strongest effect is seen in the third order correlations. The quadrature correlation between phase quadrature of the idler and the amplitude quadratures of the signal and pump becomes % \textcolor{blue}{$\chi e^{2\chi^\prime}$ ignored}
\begin{align}
& V_{abc} = \tfrac{\chi}{16 \chi^\prime}
\;\;e^{4 \chi^\prime}{-\tfrac{\chi}{4}e^{2\chi^\prime}}.
\end{align}
%\textcolor{blue}{(Blue term about 1 order of magnitude smaller than the $e^{4\chi^\prime}$ term but has a visible effect when $\alpha=10$.)} 
As this moment is linear in $\chi$ it should be the first quantum effect to become observable as we enter the pump depletion regime at high pump powers.

{\it Conclusion}: We have derived non-linear Heisenberg equations describing the evolution of quantum fields through the trilinear Hamiltonian which models parametric amplification with pump depletion. Unlike previous treatments we perform our perturbative expansion in such a way as to allow the strong pump regime to be explored.  We expect our results to be immediately useful in describing and motivating squeezing experiments in the strong pump regime. Being Heisenberg picture equations they provide good intuition about the physics and can be easily adapted to account for imperfections such as loss and excess noise.  We also expect our solutions to stimulate investigations into novel quantum protocols and technologies which may be enabled by the non-Gaussian correlations \cite{STR18} that emerge as we push further into the depleted pump regime of squeezing.

{\it Acknowledgements}: This research was supported by the Australian Research Council (ARC) under the Centre of Excellence for Quantum Computation and Communication Technology (CE170100012).

{\it Note added}: After the completion of this work we became aware of a related, but distinct approach to pump depletion in single-mode squeezing following a Schr\"{o}dinger picture approach \cite{YAN21}

\newpage

\onecolumngrid

%\section{\texorpdfstring{Full Mode Expansions to $\chi^8$}{}}\label{appx:fullResults}
\section{Appendix}
\subsection{
Full Mode Expansions to $\chi^8$}
\label{appx:fullResults}

The exact unitary describing the parametric amplification process is given by
\begin{align*}
    U = \exp{-i\chi(a^\dag b^\dag c + abc^\dag)}.
\end{align*}
Notice that the approximation that leads us back to the quadratic form in Eq.1 of the main text is the replacement $c \to \langle c \rangle = \alpha$. We want to know the full forms of operators $a, b, c$ in the Heisenberg picture, which we denote $a_o,b_o,c_o$. We no longer obtain the simple closed form linear equations of Eq.3%\ref{Hquad}
, nevertheless these Heisenberg operators 
%they 
can be evaluated to any desired order using the Baker-Campbell-Hausdorff formula. For example, the signal mode is
\begin{align*}
    a_o = e^{G} a e^{-G} & = a + [G,a] + \tfrac{1}{2!} [G,[G,a]] + \tfrac{1}{3!}[G,[G,[G,a]]] + \dots
\end{align*}
% Absorb the constant $\lambda$ and $t$ into a single coupling constant $\chi$, then 
In this case $G = i\chi (a^\dag b^\dag c + abc^\dag)$. For reference, the Heisenberg operator for the signal mode $a_o$ to order $\chi^8$ is:
\begin{align*}
    a_o & = a - i \chi b^\dag c + \tfrac{\chi^2}{2!}\qty(-a b^{\dag} b + a c^{\dag} c)  \\
    & + \tfrac{i\chi^3}{3!}\qty(2 a^{\dag} a b^{\dag} c - b^{\dag} c^{\dag} c^2 + b^{\dag} c + b^{\dag2} b c - 2 a^2 b c^{\dag})  \\
    & + \tfrac{\chi^4}{4!}\qty(4 a^{\dag} b^{\dag2} c^2 + 4 a^{\dag} a^2 b^{\dag} b - 4 a^{\dag} a^2 c^{\dag} c + a b^{\dag} b - 10 a b^{\dag} b c^{\dag} c + a b^{\dag2} b^2 - 7 a c^{\dag} c + a c^{\dag2} c^2)  \\
    & + \tfrac{i\chi^5}{5!}\bigg(28   a^{\dag} a b^{\dag} c^{\dag} c^2 - 16   a^{\dag} a b^{\dag} c - 28   a^{\dag} a b^{\dag2} b c + 8   a^{\dag} a^3 b c^{\dag} - 8   a^{\dag2} a^2 b^{\dag} c + 25   b^{\dag} c^{\dag} c^2 -   b^{\dag} c^{\dag2} c^3 -   b^{\dag} c  \\
    & \qquad + 14   b^{\dag2} b c^{\dag} c^2 - 3   b^{\dag2} b c -   b^{\dag3} b^2 c + 16   a^2 b^{\dag} b^2 c^{\dag} + 12   a^2 b c^{\dag} - 16   a^2 b c^{\dag2} c \bigg)  \\
    & + \tfrac{\chi^6}{6!}\bigg(44 a^{\dag} b^{\dag2} c^{\dag} c^3 - 72 a^{\dag} b^{\dag2} c^2 - 44 a^{\dag} b^{\dag3} b c^2 - 28 a^{\dag} a^2 b^{\dag} b + 216 a^{\dag} a^2 b^{\dag} b c^{\dag} c - 44 a^{\dag} a^2 b^{\dag2} b^2 \\
    & \qquad + 68 a^{\dag} a^2 c^{\dag} c - 44 a^{\dag} a^2 c^{\dag2} c^2 - 72 a^{\dag2} a b^{\dag2} c^2 - 16 a^{\dag2} a^3 b^{\dag} b + 16 a^{\dag2} a^3 c^{\dag} c - a b^{\dag} b + 216 a b^{\dag} b c^{\dag} c \\
    & \qquad - 91 a b^{\dag} b c^{\dag2} c^2 - 3 a b^{\dag2} b^2 + 91 a b^{\dag2} b^2 c^{\dag} c - a b^{\dag3} b^3 + 41 a c^{\dag} c - 85 a c^{\dag2} c^2 + a c^{\dag3} c^3 - 40 a^3 b^2 c^{\dag2}\bigg) \\
    & + \tfrac{i\chi^7}{7!} \bigg(-1386   a^{\dag} a b^{\dag} c^{\dag} c^2 + 270   a^{\dag} a b^{\dag} c^{\dag2} c^3 + 98   a^{\dag} a b^{\dag} c - 1204   a^{\dag} a b^{\dag2} b c^{\dag} c^2 + 598   a^{\dag} a b^{\dag2} b c \\
    & \qquad + 270   a^{\dag} a b^{\dag3} b^2 c - 416   a^{\dag} a^3 b^{\dag} b^2 c^{\dag} - 128   a^{\dag} a^3 b c^{\dag} + 416   a^{\dag} a^3 b c^{\dag2} c + 160   a^{\dag2} b^{\dag3} c^3 \\
    & \qquad - 496   a^{\dag2} a^2 b^{\dag} c^{\dag} c^2 + 144   a^{\dag2} a^2 b^{\dag} c + 496   a^{\dag2} a^2 b^{\dag2} b c - 32   a^{\dag2} a^4 b c^{\dag} + 32   a^{\dag3} a^3 b^{\dag} c \\
    & \qquad - 401   b^{\dag} c^{\dag} c^2 + 264   b^{\dag} c^{\dag2} c^3 -   b^{\dag} c^{\dag3} c^4 +   b^{\dag} c  - 602   b^{\dag2} b c^{\dag} c^2 + 135   b^{\dag2} b c^{\dag2} c^3 + 7   b^{\dag2} b c \\
    & \qquad - 135   b^{\dag3} b^2 c^{\dag} c^2 + 6   b^{\dag3} b^2 c +   b^{\dag4} b^3 c - 338   a^2 b^{\dag} b^2 c^{\dag} + 700   a^2 b^{\dag} b^2 c^{\dag2} c - 138   a^2 b^{\dag2} b^3 c^{\dag} \\
    & \qquad - 70   a^2 b c^{\dag} + 910   a^2 b c^{\dag2} c - 138   a^2 b c^{\dag3} c^2 \bigg)  \\
    & + \tfrac{\chi^8}{8!}\bigg(-4680 a^{\dag} b^{\dag2} c^{\dag} c^3 + 408 a^{\dag} b^{\dag2} c^{\dag2} c^4 + 1104 a^{\dag} b^{\dag2} c^2 - 2064 a^{\dag} b^{\dag3} b c^{\dag} c^3 + 1752 a^{\dag} b^{\dag3} b c^2 \\
    & \qquad + 408 a^{\dag} b^{\dag4} b^2 c^2 + 168 a^{\dag} a^2 b^{\dag} b - 11616 a^{\dag} a^2 b^{\dag} b c^{\dag} c + 7272 a^{\dag} a^2 b^{\dag} b c^{\dag2} c^2 + 936 a^{\dag} a^2 b^{\dag2} b^2 \\
    & \qquad - 7272 a^{\dag} a^2 b^{\dag2} b^2 c^{\dag} c + 408 a^{\dag} a^2 b^{\dag3} b^3 - 840 a^{\dag} a^2 c^{\dag} c + 4536 a^{\dag} a^2 c^{\dag2} c^2 - 408 a^{\dag} a^2 c^{\dag3} c^3\\
    & \qquad + 896 a^{\dag} a^4 b^2 c^{\dag2} - 3216 a^{\dag2} a b^{\dag2} c^{\dag} c^3 + 3264 a^{\dag2} a b^{\dag2} c^2 + 3216 a^{\dag2} a b^{\dag3} b c^2 + 272 a^{\dag2} a^3 b^{\dag} b \\
    & \qquad - 3872 a^{\dag2} a^3 b^{\dag} b c^{\dag} c + 912 a^{\dag2} a^3 b^{\dag2} b^2 - 496 a^{\dag2} a^3 c^{\dag} c + 912 a^{\dag2} a^3 c^{\dag2} c^2 + 1088 a^{\dag3} a^2 b^{\dag2} c^2 \\
    & \qquad + 64 a^{\dag3} a^4 b^{\dag} b - 64 a^{\dag3} a^4 c^{\dag} c + a b^{\dag} b - 3602 a b^{\dag} b c^{\dag} c + 10410 a b^{\dag} b c^{\dag2} c^2 - 820 a b^{\dag} b c^{\dag3} c^3 \\
    & \qquad + 7 a b^{\dag2} b^2 - 4134 a b^{\dag2} b^2 c^{\dag} c + 3414 a b^{\dag2} b^2 c^{\dag2} c^2 + 6 a b^{\dag3} b^3 - 820 a b^{\dag3} b^3 c^{\dag} c + a b^{\dag4} b^4 \\
    & \qquad - 239 a c^{\dag} c + 3607 a c^{\dag2} c^2 - 810 a c^{\dag3} c^3 + a c^{\dag4} c^4 + 1392 a^3 b^{\dag} b^3 c^{\dag2} + 1792 a^3 b^2 c^{\dag2} - 1392 a^3 b^2 c^{\dag3} c \bigg) \\
    & + \dots
\end{align*}
And the time-evolved operator for the idler mode $b_o$ can be found by swapping $a$ and $b$ in the above formula. The formula for the pump mode is 
\begin{align*}
    c_o & = c - i\chi ab + \tfrac{\chi^2}{2}\qty(-a^{\dag} a c - b^{\dag} b c - c) + \tfrac{i\chi^3}{3!} \qty(2   a^{\dag} b^{\dag} c^2 +   a^{\dag} a^2 b +   a b^{\dag} b^2 +   a b - 2   a b c^{\dag} c)  \\
    & + \tfrac{\chi^4}{4!}\qty(10 a^{\dag} a b^{\dag} b c - 4 a^{\dag} a c^{\dag} c^2 + 3 a^{\dag} a c + a^{\dag2} a^2 c - 4 b^{\dag} b c^{\dag} c^2 + 3 b^{\dag} b c + b^{\dag2} b^2 c - 4 c^{\dag} c^2 - 4 a^2 b^2 c^{\dag} + c)  \\
    & + \tfrac{i\chi^5}{5!}\bigg(8   a^{\dag} b^{\dag} c^{\dag} c^3 - 20   a^{\dag} b^{\dag} c^2 - 16   a^{\dag} b^{\dag2} b c^2 - 14   a^{\dag} a^2 b^{\dag} b^2 - 3   a^{\dag} a^2 b + 28   a^{\dag} a^2 b c^{\dag} c - 16   a^{\dag2} a b^{\dag} c^2 \\
    & \qquad -   a^{\dag2} a^3 b - 3   a b^{\dag} b^2 + 28   a b^{\dag} b^2 c^{\dag} c -   a b^{\dag2} b^3 -   a b + 40   a b c^{\dag} c - 8   a b c^{\dag2} c^2 \bigg)  \\
    & + \tfrac{\chi^6}{6!}\bigg(216 a^{\dag} a b^{\dag} b c^{\dag} c^2 - 148 a^{\dag} a b^{\dag} b c - 91 a^{\dag} a b^{\dag2} b^2 c + 148 a^{\dag} a c^{\dag} c^2 - 16 a^{\dag} a c^{\dag2} c^3 - 7 a^{\dag} a c + 44 a^{\dag} a^3 b^2 c^{\dag} \\ 
    & \qquad - 40 a^{\dag2} b^{\dag2} c^3 - 91 a^{\dag2} a^2 b^{\dag} b c + 44 a^{\dag2} a^2 c^{\dag} c^2 - 6 a^{\dag2} a^2 c - a^{\dag3} a^3 c + 148 b^{\dag} b c^{\dag} c^2 - 16 b^{\dag} b c^{\dag2} c^3 - 7 b^{\dag} b c \\
    & \qquad + 44 b^{\dag2} b^2 c^{\dag} c^2 - 6 b^{\dag2} b^2 c - b^{\dag3} b^3 c + 60 c^{\dag} c^2 - 16 c^{\dag2} c^3 + 44 a^2 b^{\dag} b^3 c^{\dag} + 60 a^2 b^2 c^{\dag} - 72 a^2 b^2 c^{\dag2} c - c \bigg)  \\
    & + \tfrac{i\chi^7}{7!}\bigg(-704   a^{\dag} b^{\dag} c^{\dag} c^3 + 32   a^{\dag} b^{\dag} c^{\dag2} c^4 + 222   a^{\dag} b^{\dag} c^2 - 416   a^{\dag} b^{\dag2} b c^{\dag} c^3 + 490   a^{\dag} b^{\dag2} b c^2 + 138   a^{\dag} b^{\dag3} b^2 c^2 \\
    & \qquad + 208   a^{\dag} a^2 b^{\dag} b^2 - 1204   a^{\dag} a^2 b^{\dag} b^2 c^{\dag} c + 135   a^{\dag} a^2 b^{\dag2} b^3 + 7   a^{\dag} a^2 b - 1022   a^{\dag} a^2 b c^{\dag} c + 496   a^{\dag} a^2 b c^{\dag2} c^2 \\
    & \qquad - 416   a^{\dag2} a b^{\dag} c^{\dag} c^3 + 490   a^{\dag2} a b^{\dag} c^2 + 700   a^{\dag2} a b^{\dag2} b c^2 + 135   a^{\dag2} a^3 b^{\dag} b^2 + 6   a^{\dag2} a^3 b - 270   a^{\dag2} a^3 b c^{\dag} c \\
    & \qquad + 138   a^{\dag3} a^2 b^{\dag} c^2 +   a^{\dag3} a^4 b + 7   a b^{\dag} b^2 - 1022   a b^{\dag} b^2 c^{\dag} c + 496   a b^{\dag} b^2 c^{\dag2} c^2 + 6   a b^{\dag2} b^3 - 270   a b^{\dag2} b^3 c^{\dag} c \\
    & \qquad +   a b^{\dag3} b^4 +   a b - 522   a b c^{\dag} c + 848   a b c^{\dag2} c^2 - 32   a b c^{\dag3} c^3 + 160   a^3 b^3 c^{\dag2} \bigg)  \\
    & + \tfrac{\chi^8}{8!}\bigg(-17472 a^{\dag} a b^{\dag} b c^{\dag} c^2 + 3872 a^{\dag} a b^{\dag} b c^{\dag2} c^3 + 1826 a^{\dag} a b^{\dag} b c - 7272 a^{\dag} a b^{\dag2} b^2 c^{\dag} c^2 + 3246 a^{\dag} a b^{\dag2} b^2 c \\ 
    & \qquad + 820 a^{\dag} a b^{\dag3} b^3 c - 3768 a^{\dag} a c^{\dag} c^2 + 3376 a^{\dag} a c^{\dag2} c^3 - 64 a^{\dag} a c^{\dag3} c^4 + 15 a^{\dag} a c - 2064 a^{\dag} a^3 b^{\dag} b^3 c^{\dag} \\
    & \qquad - 1512 a^{\dag} a^3 b^2 c^{\dag} + 3216 a^{\dag} a^3 b^2 c^{\dag2} c - 896 a^{\dag2} b^{\dag2} c^{\dag} c^4 + 2384 a^{\dag2} b^{\dag2} c^3 + 1392 a^{\dag2} b^{\dag3} b c^3 \\
    & \qquad - 7272 a^{\dag2} a^2 b^{\dag} b c^{\dag} c^2 + 3246 a^{\dag2} a^2 b^{\dag} b c + 3414 a^{\dag2} a^2 b^{\dag2} b^2 c - 2736 a^{\dag2} a^2 c^{\dag} c^2 + 912 a^{\dag2} a^2 c^{\dag2} c^3 \\
    &\qquad + 25 a^{\dag2} a^2 c - 408 a^{\dag2} a^4 b^2 c^{\dag} + 1392 a^{\dag3} a b^{\dag2} c^3 + 820 a^{\dag3} a^3 b^{\dag} b c - 408 a^{\dag3} a^3 c^{\dag} c^2 + 10 a^{\dag3} a^3 c \\
    & \qquad + a^{\dag4} a^4 c - 3768 b^{\dag} b c^{\dag} c^2 + 3376 b^{\dag} b c^{\dag2} c^3 - 64 b^{\dag} b c^{\dag3} c^4 + 15 b^{\dag} b c - 2736 b^{\dag2} b^2 c^{\dag} c^2 + 912 b^{\dag2} b^2 c^{\dag2} c^3 \\
    &\qquad + 25 b^{\dag2} b^2 c - 408 b^{\dag3} b^3 c^{\dag} c^2 + 10 b^{\dag3} b^3 c + b^{\dag4} b^4 c - 744 c^{\dag} c^2 + 1552 c^{\dag2} c^3 - 64 c^{\dag3} c^4 - 1512 a^2 b^{\dag} b^3 c^{\dag} \\
    &\qquad + 3216 a^2 b^{\dag} b^3 c^{\dag2} c - 408 a^2 b^{\dag2} b^4 c^{\dag} - 744 a^2 b^2 c^{\dag} + 6384 a^2 b^2 c^{\dag2} c - 1088 a^2 b^2 c^{\dag3} c^2 + c \bigg) \\
    & + \dots
\end{align*}

%\section{\texorpdfstring{Full Output Operators to $\mathcal O(\chi^2)$}{}}\label{appx:orderChi2Ops}
\subsection{Full Output Operators to $\mathcal O(\chi^2)$}\label{appx:orderChi2Ops}

We perform a $c = \alpha + \delta c$ expansion, and retain only the terms of the form $\alpha^n\chi^n$ (assumed to be of order 1) and $\alpha^{n-1}\chi^n$ (of $\mathcal O(\chi)$). For example, with the pump mode $c_o$, only a small number of terms could potentially contribute at $\mathcal O(\chi)$:
\begin{align*}
    c_o & = c-i\chi ab + \tfrac{\chi^2}{2}(-c-a^\dag ac-b^\dag bc) \\
    & + \tfrac{i\chi^3}{3!}(2a^\dag b^\dag c^2 - 2abc^\dag c) +\tfrac{\chi^4}{4!}(-4c^\dag c^2-4a^\dag ac^\dag c^2
    - 4b^\dag bc^\dag c^2  ) \\ 
    & + \tfrac{i\chi^5}{5!}(8a^\dag b^\dag c^\dag c^3 -8abc^{\dag2}c^2)  + \tfrac{\chi^6}{6!}(-16c^{\dag2}c^3-16a^\dag ac^{\dag2}c^3-16b^\dag bc^{\dag2}c^3) \\
    & + \tfrac{i\chi^7}{7!}(32a^\dag b^\dag c^{\dag2}c^4-32abc^{\dag3}c^3) + \tfrac{\chi^8}{8!}(-64c^{\dag3}c^4-64a^\dag ac^{\dag3}c^4-64b^\dag bc^{\dag3}c^4) + \dots
\end{align*}
Now substitute $c=\alpha+\delta c$ (and therefore $c^\dag = \alpha + \delta c^\dag)$, keeping only $\mathcal O(\chi)$ terms, we have
\begin{align*}
    c_o & = \ \alpha + \delta c - i\chi ab + \tfrac{\chi^2}{2!}\qty(- \alpha -  \alpha  a^{\dag}  a -  \alpha  b^{\dag}  b) \\
    & + \tfrac{i\chi^3}{3!} \qty(2      \alpha ^ 2  a^{\dag}  b^{\dag} - 2      \alpha ^ 2  a  b) + \tfrac{\chi^4}{4!} \qty(-4   \alpha ^ 3 - 4   \alpha ^ 3  a^{\dag}  a - 4   \alpha ^ 3  b^{\dag}  b) \\
    & + \tfrac{i\chi^5}{5!} \qty(8      \alpha ^ 4  a^{\dag}  b^{\dag} - 8      \alpha ^ 4  a  b) + \tfrac{\chi^6}{6!} \qty(-16   \alpha ^ 5 - 16   \alpha ^ 5  a^{\dag}  a - 16   \alpha ^ 5  b^{\dag}  b) \\
    & + \tfrac{i\chi^7}{7!}\qty(32      \alpha ^ 6  a^{\dag}  b^{\dag} - 32      \alpha ^ 6  a  b) + \tfrac{\chi^8}{8!} \qty(-64   \alpha ^ 7 - 64   \alpha ^ 7  a^{\dag}  a - 64   \alpha ^ 7  b^{\dag}  b) + \dots 
    \end{align*}
Group the terms according to the operators they are multiplied to. 
\begin{align*}
    \Rightarrow c_o & = \alpha - ( \tfrac{\alpha\chi^2}{2!} + \tfrac{4\alpha^3\chi^4}{4!} + \tfrac{16\alpha^5\chi^6}{6!} + \tfrac{64\alpha^7\chi^8}{8!} + \dots ) \\
    & + \delta c - (a^\dag a + b^\dag b) (\tfrac{\alpha\chi^2}{2!} + \tfrac{4\alpha^3\chi^4}{4!} + \tfrac{16\alpha^5\chi^6}{6!} + \tfrac{64\alpha^7\chi^8}{8!} + \dots) \\
    & + ia^\dag b^\dag ( \tfrac{2\alpha^2\chi^3}{3!} + \tfrac{8\alpha^4\chi^5}{5!} + \tfrac{32\alpha^6\chi^7}{7!} + \dots ) \\
    & -iab (\chi + \tfrac{2\alpha^2\chi^3}{3!} + \tfrac{8\alpha^4\chi^5}{5!} + \tfrac{32\alpha^6\chi^7}{7!} + \dots).
\end{align*}
Note just as we decomposed $c = \alpha + \delta c$, we can decompose the Heisenberg operator into an amplitude part and a noise part, $c_o = \alpha_o + \delta c_o$, both parts are time-dependent. We see from above that the first line of $c_o$ consists only of pure numbers and no operators, and is therefore the amplitude $\alpha_o$; anything on the second line and below are the noise part $\delta c_o$. 

There are clear patterns to the first few terms of each infinite series appeared above. Assuming the patterns persist indefinitely (checked to order $\alpha^{14}\chi^{15}$), we may express each infinite series as a sum, and use Mathematica to find the closed forms of these series:
\begin{align*}
    \tfrac{\alpha\chi^2}{2!} + \tfrac{4\alpha^3\chi^4}{4!} + \tfrac{16\alpha^5\chi^6}{6!} + \dots & = \sum_{n=1}^\infty \tfrac{\alpha^{2n-1}\chi^{2n}\cdot 4^{n-1}}{(2n)!} %= -\tfrac{1}{4\alpha} + \tfrac{1}{4\alpha}\cosh 2\alpha\chi 
    = \tfrac{\chi}{2\chi^\prime}\sinh^2\chi^\prime, \\
    \tfrac{2\alpha^2\chi^3}{3!} + \tfrac{8\alpha^4\chi^5}{5!} + \tfrac{32\alpha^6\chi^7}{7!} + \dots &= \sum_{n=1}^\infty \tfrac{\alpha^{2n}\chi^{2n+1}\cdot 2 \cdot 4^{n-1}}{(2n+1)!}% = -\tfrac{\chi}{2} + \tfrac{1}{4\alpha}\sinh 2\alpha\chi 
    = -\tfrac{\chi}{2} + \tfrac{\chi}{2\chi^\prime}\sinh\chi^\prime\cosh\chi^\prime, \\
    \chi + \tfrac{2\alpha^2\chi^3}{3!} + \tfrac{8\alpha^4\chi^5}{5!} + \tfrac{32\alpha^6\chi^7}{7!} + \dots & = %\chi -\tfrac{\chi}{2} + \tfrac{\chi}{2\chi^\prime}\sinh\chi^\prime\cosh\chi^\prime =
    \tfrac{\chi}{2} + \tfrac{\chi}{2\chi^\prime}\sinh\chi^\prime\cosh\chi^\prime,
\end{align*}
where we %used the identities $\cosh 2x = \sinh^2x + \cosh^2x = 2\sinh^2x + 1$ and $\sinh2x = 2\sinh x \cosh x$, and 
again defined $\chi^\prime\equiv \alpha\chi$. So $c_o$ is 
\begin{align*}
\begin{split}
    c_o & = \alpha_o + \delta c_o \\
    & = \alpha - \tfrac{\chi}{2\chi^\prime}\sinh^2\chi^\prime + \delta c  - (a^\dag a + b^\dag b) \tfrac{\chi}{2\chi^\prime}\sinh^2\chi^\prime \\
    & - ia^\dag b^\dag \tfrac{\chi}{2}(1 - \tfrac{1}{\chi^\prime}\sinh\chi^\prime \cosh\chi^\prime) - iab \tfrac{\chi}{2} (1 + \tfrac{1}{\chi^\prime}\sinh\chi^\prime\cosh\chi^\prime),
\end{split}
\end{align*}
where we see that
\begin{align*}
    \alpha_o & = \alpha - \tfrac{\chi}{2\chi^\prime}\sinh^2\chi^\prime, \\
    \text{and } \ \delta c_o & = \delta c  - (a^\dag a + b^\dag b) \tfrac{\chi}{2\chi^\prime}\sinh^2\chi^\prime - ia^\dag b^\dag \tfrac{\chi}{2}(1 - \tfrac{1}{\chi^\prime}\sinh\chi^\prime \cosh\chi^\prime) - iab \tfrac{\chi}{2} (1 + \tfrac{1}{\chi^\prime}\sinh\chi^\prime\cosh\chi^\prime).
\end{align*}
\indent For a fully self-consistent model capable of calculating non-trivial expectation values, we need to include the $\mathcal O(\chi^2)$ terms in the output modes $a_o, b_o, c_o$ as well. That is, after the $c=\alpha+\delta c$ expansion, on top of the $\alpha^n\chi^n$ and $\alpha^{n-1}\chi^n$ terms, we now also retain terms of the form $\alpha^{n-2}\chi^n$. This introduces many additional infinite series. For illustrative purposes, let's focus on a couple of them. In $c_o$, consider terms proportional to $a^\dag a c^{\dag n}c^m$, the terms that could contribute are
\begin{align*}
    c_o & = \dots - \tfrac{\chi^2}{2}a^\dag ac - \tfrac{\chi^4}{4!}4a^\dag ac^\dag c^2 - \tfrac{\chi^6}{6!}16a^\dag ac^{\dag2}c^3 - \tfrac{\chi^8}{8!}64a^\dag ac^{\dag3}c^4 + \dots,
\end{align*}
now perform the $c=\alpha+\delta c$ expansion, we have
\begin{align*}
    c_o & = \dots - a^\dag a (\tfrac{\chi^2}{2}(\alpha+\delta c) + \tfrac{\chi^4}{4!}4(\alpha+\delta c^\dag)(\alpha+\delta c)^2 + \tfrac{\chi^6}{6!}16(\alpha+\delta c^\dag)^2(\alpha+\delta c)^3 + \tfrac{\chi^8}{8!}64(\alpha+\delta c^\dag)^3(\alpha+\delta c)^4 + \dots) \\
    & = \dots - a^\dag a (\tfrac{\chi^2}{2}\alpha + \tfrac{4\chi^4}{4!}\alpha^3 + \tfrac{16\chi^6}{6!}\alpha^5 + \tfrac{64\chi^8}{8!}\alpha^7 + \dots) \\
    & \qquad \ \, - a^\dag a(\tfrac{\chi^2}{2}\delta c + \tfrac{8\chi^4}{4!}\alpha^2\delta c+ \tfrac{48\chi^6}{6!}\alpha^4\delta c + \tfrac{256\chi^8}{8!}\alpha^6\delta c + \dots ) \\
    & \qquad \ \, - a^\dag a(\tfrac{4\chi^4}{4!}\alpha^2\delta c^\dag  + \tfrac{32\chi^6}{6!}\alpha^4\delta c^\dag + \tfrac{192\chi^8}{8!}\alpha^6\delta c^\dag + \dots),
\end{align*}
giving three infinite series. The first one we've already seen, it is of $\mathcal O(\chi)$ and equals to $-a^\dag a\tfrac{\chi}{2\chi^\prime}\sinh^2\chi^\prime$. The last two series are new, and can be put into closed form expressions as
\begin{align*}
    c_o & = \dots - a^\dag a \delta c\, \chi^2 \sum_{n=0}^\infty 2^{2n-1}(2n+2)\cdot \frac{\chi^{\prime 2n}}{(2n+2)!}  - a^\dag a \delta c^\dag \,\chi^2 \sum_{n=1}^\infty 2^{2n}n\frac{\chi^{\prime2n}}{(2n+2)!} \\
    & = \dots -a^\dag a\delta c \,\chi^2 \tfrac{1-\cosh2\chi^\prime+2\sinh^2\chi^\prime+\chi^\prime\sinh2\chi^\prime}{4\chi^{\prime2}} - a^\dag a \delta c^\dag \,\chi^2 \tfrac{1-\cosh2\chi^\prime+\chi^\prime\sinh2\chi^\prime}{4\chi^{\prime2}}.
\end{align*}
With some work, all second order terms can be grouped into series which can then be expressed as closed form expressions like the ones above. 

% There will be many more additional infinite series of $\mathcal O(\chi^2)$. For example, one of them is
% \begin{align*}
%     c_o & = \dots - \delta c( \tfrac{\chi^2}{2} + \tfrac{\alpha^2\chi^4}{3} + \tfrac{\alpha^4\chi^6}{15} + \tfrac{2\alpha^6\chi^8}{315} + \dots) \\
%     & = \dots  - \delta c \,\chi^2 \sum_{n=0}^\infty 2^{2n-1}(2n+2)\cdot \tfrac{\chi^{\prime2n}}{(2n+2)!} \\
%     & = \dots- \delta c \, \chi^2  \tfrac{1-\cosh2\chi^\prime+2\sinh^2\chi^\prime+\chi^\prime\sinh2\chi^\prime}{4\chi^{\prime2}}.
% \end{align*}
\indent We now simply list the final results. We find that the signal mode $a_o$ to order $\chi^2$ is 
\begin{align}
    a_o & = a \qty[\cosh\chi^\prime + (\delta c + \delta c^\dag) \tfrac{\chi}{2}\sinh\chi^\prime] - ib^\dag \big[\sinh \chi^\prime + \tfrac{\chi}{2}\cosh\chi^\prime(\delta c + \delta c^\dag)+ \tfrac{\chi}{2\chi^\prime} \sinh\chi^\prime(\delta c - \delta c^\dag)\big] \nonumber \\
    & + \chi^2 \bigg( A_a a + A_{b^\dag} b^\dag+ A_{a^2b} a^2b + A_{ab^\dag b} ab^\dag b +A_{a^\dag a b^\dag} a^\dag a b^\dag  + A_{a^\dag b^{\dag2}} a^\dag b^{\dag2} + A_{a^\dag a^2}a^\dag a^2 + A_{a\delta c^2}a\delta c^2  \nonumber \\
    & + A_{a\delta c^{\dag2}} a\delta c^{\dag2} + A_{a\delta c^\dag \delta c} a\delta c^\dag \delta c + A_{b^{\dag2}b}b^{\dag2}b + A_{b^\dag\delta c^2}b^\dag\delta c^2 + A_{b^\dag \delta c^\dag \delta c} b^\dag \delta c^\dag \delta c + A_{b^\dag \delta c^{\dag2}} b^\dag \delta c^{\dag2} \bigg), \label{eqn:signalFULL}
\end{align}
where
\begin{align*}
    A_{a} & = -\sum_{n=2}^\infty \qty(\tfrac{9^n-1}{8}-\tfrac{n(n+1)}{2})\cdot \tfrac{\chi^{\prime2n-2}}{(2n)!} = -\tfrac{-\cosh\chi^\prime-\chi^{\prime2}\cosh\chi^\prime+\cosh3\chi^\prime-3\chi^\prime\sinh\chi^\prime}{8\chi^{\prime2}}, \\
    A_{b^\dag} & = i\sum_{n=1}^\infty Y(n)\tfrac{\chi^{\prime2n-1}}{(2n+1)!} = i\cdot \tfrac{-5\chi^\prime\cosh\chi^\prime+2\sinh\chi^\prime-\chi^{\prime2}\sinh\chi^\prime+\sinh3\chi^\prime}{8\chi^{\prime2}},\\
    A_{a^2b} & = -i\sum_{n=1}^\infty(a(n)+n)\cdot \tfrac{\chi^{\prime2n-1}}{(2n+1)!} = -i\cdot \tfrac{4\chi^\prime\cosh\chi^\prime-7\sinh\chi^\prime+\sinh3\chi^\prime}{16\chi^{\prime2}}, \\
    A_{ab^\dag b} & = -\sum_{n=1}^\infty \tfrac{9^n-1}{8}\cdot \tfrac{\chi^{\prime2n-2}}{(2n)!} = \tfrac{\cosh\chi^\prime-\cosh3\chi^\prime}{8\chi^{\prime2}}, \\
    A_{a^\dag ab^\dag} & = i\sum_{n=1}^\infty (Y(n)+\tfrac{n(n+1)}{2})\cdot \tfrac{\chi^{\prime2n-1}}{(2n+1)!} = i\cdot \tfrac{-4\chi^\prime\cosh\chi^\prime+\sinh\chi^\prime+\sinh3\chi^\prime}{8\chi^{\prime2}}, \\
    A_{a^\dag b^{\dag2}} & = - A_{a^\dag a^2} = \sum_{n=2}^\infty (\tfrac{9^n-1}{16}-\tfrac{n}{2})\cdot \tfrac{\chi^{\prime2n-2}}{(2n)!} = \tfrac{-\cosh\chi^\prime+\cosh3\chi^\prime-4\chi^\prime\sinh\chi^\prime}{16\chi^{\prime2}}, \\
    A_{a\delta c^2} & = A_{a\delta c^{\dag2}} = \sum_{n=1}^\infty \tfrac{n(n+1)}{2}\cdot \tfrac{\chi^{\prime2n}}{(2n+2)!} = \tfrac{\chi^\prime\cosh\chi^\prime-\sinh\chi^\prime}{8\chi^\prime}, \\
    A_{a\delta c^\dag \delta c} & = \sum_{n=0}^\infty(n+1)^2\tfrac{\chi^{\prime2n}}{(2n+2)!} = \tfrac{\chi^\prime\cosh\chi^\prime+\sinh\chi^\prime}{4\chi^\prime}, \\
    A_{b^{\dag2}b} & = i\sum_{n=1}^\infty a(n)\cdot \tfrac{\chi^{\prime2n-1}}{(2n+1)!}=i\cdot \tfrac{-4\chi^\prime\cosh\chi^\prime+\sinh\chi^\prime+\sinh3\chi^\prime}{16\chi^{\prime2}},\\
    A_{b^\dag\delta c^2} & =  \frac{1}{2}A_{b^\dag \delta c^\dag \delta c} = -i\sum_{n=1}^\infty \tfrac{n(n+1)}{2}\cdot \tfrac{\chi^{\prime2n-1}}{(2n+1)!} = -i\cdot \tfrac{\chi^\prime\cosh\chi^\prime-\sinh\chi^\prime+\chi^{\prime2}\sinh\chi^\prime}{8\chi^\prime2}, \\
    A_{b^\dag \delta c^{\dag2}} & =-i \sum_{n=1}^\infty (\tfrac{n(n+1)}{2})\cdot \tfrac{\chi^{\prime2n+1}}{(2n+3)!} = -i\cdot \tfrac{-3\chi^\prime\cosh\chi^\prime + 3\sinh\chi^\prime + \chi^{\prime2}\sinh\chi^\prime}{8\chi^{\prime2}}.
\end{align*}
And for the pump mode:
\begin{align}
    c_o & = \alpha - \tfrac{\chi}{2\chi^\prime}\sinh^2\chi^\prime + \chi^3 C_\alpha  \nonumber \\
    & + \delta c  - (a^\dag a + b^\dag b) \tfrac{\chi}{2\chi^\prime}\sinh^2\chi^\prime - ia^\dag b^\dag \tfrac{\chi}{2}(1 - \tfrac{1}{\chi^\prime}\sinh\chi^\prime \cosh\chi^\prime) - iab \tfrac{\chi}{2} (1 + \tfrac{1}{\chi^\prime}\sinh\chi^\prime\cosh\chi^\prime) \nonumber \\
    & + \chi^2 \bigg( C_{\delta c} \delta c + C_{a^\dag a\delta c}a^\dag a \delta c + C_{b^\dag b\delta c} b^\dag b\delta c + C_{ab\delta c} ab\delta c + C_{ab\delta c^\dag}ab\delta c^\dag +C_{a^\dag b^\dag \delta c} a^\dag b^\dag \delta c \nonumber \\
    & + C_{\delta c^\dag} \delta c^\dag + C_{a^\dag a\delta c^\dag}a^\dag a\delta c^\dag + C_{b^\dag b\delta c^\dag} b^\dag b\delta c^\dag + C_{a^\dag b^\dag \delta c^\dag}a^\dag b^\dag \delta c^\dag\bigg), \label{eqn:pumpFULL}
\end{align}
where
\begin{align*}
    C_{\alpha} & = \sum_{n=1}^\infty Z(n)\cdot \tfrac{\chi^{\prime2n-1}}{(2n+2)!} = \tfrac{-3-4\chi^{\prime2}+(2-4\chi^{\prime2})\cosh2\chi^\prime + \cosh4\chi^\prime-2\chi^\prime\sinh2\chi^\prime}{32\chi^{\prime3}} \\
    C_{\delta c} & = C_{a^\dag a\delta c} = C_{b^\dag b\delta c} = -\sum_{n=0}^\infty 2^{2n-1}(2n+2)\cdot \tfrac{\chi^{\prime2n}}{(2n+2)!} =- \tfrac{1-\cosh2\chi^\prime+2\sinh^2\chi^\prime+\chi^\prime\sinh2\chi^\prime}{4\chi^{\prime2}} , \\
    C_{ab\delta c} & = C_{ab\delta c^\dag} = -i\sum_{n=1}^\infty 2^{2n-2}(2n)\cdot\tfrac{\chi^{\prime2n-1}}{(2n+1)!} =-i \tfrac{2\chi^\prime\cosh2\chi^\prime-\sinh2\chi^\prime}{8\chi^{\prime2}} , \\
    C_{a^\dag b^\dag \delta c} & = i\sum_{n=2}^\infty \tfrac{2^{2n-2}n}{2} \cdot \tfrac{\chi^{\prime2n-3}}{(2n-1)!} = i\tfrac{-4\chi^\prime+2\chi^\prime\cosh2\chi^\prime+\sinh2\chi^\prime}{8\chi^{\prime2}}, \\
    C_{\delta c^\dag} & = C_{a^\dag a\delta c^\dag} = C_{b^\dag b\delta c^\dag}= -\sum_{n=1}^\infty 2^{2n}n \tfrac{\chi^{\prime2n}}{(2n+2)!} = - \tfrac{1-\cosh2\chi^\prime+\chi^\prime\sinh2\chi^\prime}{4\chi^{\prime2}}  , \\
    C_{a^\dag b^\dag \delta c^\dag} & = i\sum_{n=1}^\infty 2^{2n}(2n)\cdot \tfrac{\chi^{\prime2n+1}}{(2n+3)!} = i\tfrac{4\chi^\prime+2\chi^\prime\cosh2\chi^\prime-3\sinh2\chi^\prime}{8\chi^{\prime2}}.
\end{align*}
In the above, the expression $a(n), X(n), Y(n)$ and $Z(n)$ are given by
\begin{align*}
    a(n) &= \tfrac{3^{2n+1}-8n-3}{16}. \\
    X(n) & = 54a(n-1) + 25n - 18 - \tfrac{n(n-1)}{2}, \\
    Y(n) & = 18a(n-1) + 7n - 6 - \tfrac{n(n-1)}{2}, \\
    \text{and } \quad Z(n) &= (2n+1)!\sum_{k=0}^n \tfrac{1}{(2k)!}\tfrac{1}{[2(n-k)+1]!}\qty[X(k-1)+Y(n-k)-k(n-k)],
\end{align*}
All $n$ and $k$'s are integers. The first few numbers in each sequence are listed in Table I
\begin{table}%[H]
\centering
\begin{tabular}{c|c|c|c|c}
     $n$ & $a(n)$ & $X(n)$ & $Y(n)$ & $Z(n)$  \\ \hline
     $-1$ & $1/3$ & 0 & $2/3$ & --- \\ \hline
     0 &0 & 0  &0  & 0 \\ \hline 
     1 &1 &7  & 1&1 \\ \hline 
     2 &14 &85 &25 &60  \\ \hline
     3 &135 &810 &264 &1552    \\ \hline 
     4 &1228 &7366 &2446  &29632   \\ \hline 
     5 &11069 &66409 &22123  & 506112   \\ \hline 
     6 &99642 &597843 &199263 & 8289280  \\ \hline 
\end{tabular}
\caption{}
\end{table}

One can check that the operators are physical in the sense that the commutation relations are satisfied to $\mathcal O(\chi^2)$, namely:
\begin{align*}
    [a_o,a_o^\dag] = [b_o,b_o^\dag] = [c_o,c_o^\dag] &= 1 + \mathcal O(\chi^3), \\
    \text{all other commutation relations} &= 0 + \mathcal O(\chi^3).
\end{align*}

\indent The vast majority of the terms in the output modes do not contribute to the expectation values at $\mathcal O(\chi^2)$. Specifically, it turns out the only second order term in $a_o$ that contributes to $\expval*{a_o^\dag a_o}$ is $\chi^2 A_{b^\dag}b^\dag$; the only second order terms that contribute to $\expval*{a_ob_o}$ are $\chi^2(A_{b^\dag}b^\dag + A_aa)$. Similarly, the only second order term in $c_o$ that contributes to $\expval*{\delta c_o\delta c_o}$ is $\chi^2C_{\delta c^\dag}\delta c^\dag$, and the only second order or above term that contributes to $\expval*{c_o^\dag c_o}$ is $\chi^3 C_\alpha $. All other terms either annihilate $\bra0$ or $\ket0$ to give 0 contributions, or they only contribute to $\mathcal O(\chi^3)$ terms. So for the purpose of calculating expectation values, we may simply take
\begin{align}
    a_o & = a \qty[\cosh\chi^\prime + (\delta c + \delta c^\dag) \tfrac{\chi}{2}\sinh\chi^\prime] \nonumber \\
    & - ib^\dag \big[\sinh \chi^\prime + \tfrac{\chi}{2}\cosh\chi^\prime(\delta c + \delta c^\dag) \nonumber \\
    & \qquad + \tfrac{\chi}{2\chi^\prime} \sinh\chi^\prime(\delta c - \delta c^\dag) \big] \nonumber \\
    & \qquad + \chi^2( A_a a + A_{b^\dag} b^\dag), \label{eqn:aoappx} \\
    b_o & = b \qty[\cosh\chi^\prime + (\delta c + \delta c^\dag) \tfrac{\chi}{2}\sinh\chi^\prime] \nonumber \\
    & - ia^\dag \big[\sinh \chi^\prime + \tfrac{\chi}{2}\cosh\chi^\prime(\delta c + \delta c^\dag) \nonumber \\
    & \qquad + \tfrac{\chi}{2\chi^\prime} \sinh\chi^\prime(\delta c - \delta c^\dag) \big] \nonumber \\
    & \qquad + \chi^2( A_a b + A_{b^\dag} a^\dag), \label{eqn:boappx} \\
    c_o &=\alpha_o + \delta c_o \nonumber \\
    & = \alpha - \tfrac{\chi}{2\chi^\prime}\sinh^2\chi^\prime + \chi^3 C_\alpha \nonumber\\
    & \quad + \delta c  - (a^\dag a + b^\dag b) \tfrac{\chi}{2\chi^\prime}\sinh^2\chi^\prime \nonumber\\
    & \quad - ia^\dag b^\dag \tfrac{\chi}{2}(1 - \tfrac{1}{\chi^\prime}\sinh\chi^\prime \cosh\chi^\prime) \nonumber\\
    & \quad - iab \tfrac{\chi}{2} (1 + \tfrac{1}{\chi^\prime}\sinh\chi^\prime\cosh\chi^\prime) \nonumber \\
    & \quad + \chi^2 C_{\delta c^\dag}\delta c^\dag, \label{eqn:coappx}
\end{align}
These are `effective operators' in the sense that they give the correct results for $\expval*{a_o^\dag a_o}, \expval*{a_ob_o}, \expval*{\delta c_o^\dag \delta c_o}, \expval*{\delta c_o\delta c_o}$ and $\alpha_o^2$, and therefore all variances calculated from these operators are correct. As such, one can also check that energy is conserved,
\begin{align*}
\alpha^2 = \expval*{c_o^\dag c_o} + \expval*{a_o^\dag a_o} = \alpha_o^2 + \expval*{\delta c_o^\dag \delta c_o} + \expval*{a_o^\dag a_o}.
\end{align*}
Although not obvious, it turns out the three effective operators above also give the correct formulae for $\expval*{ab\delta c}$ and $\expval*{ab\delta c^\dag}$ up to $\mathcal O(\chi^2)$. 

The only drawback of using the effective operators is that the commutation relations given by these operator are only correct to $\mathcal O(\chi)$, not the desired $\mathcal O(\chi^2)$. Therefore, we should not use them to calculate non-normally ordered operator products, for example, $\expval*{a_oa_o^\dag}$ or $\expval*{\delta c_o\delta c_o^\dag}$.  Instead we should first normal order them, such that for example, $\expval*{a_oa_o^\dag} \to \expval*{1+a_o^\dag a_o}$ and $\expval*{\delta c_o\delta c_o^\dag} \to \expval*{1+ \delta c_o^\dag \delta c_o}$, before evaluating them.

The bottom line is, we may use equation \eqref{eqn:aoappx} to \eqref{eqn:coappx} to calculate any second- or third-order correlations to $\mathcal O(\chi^2)$, provided the correlations are normal ordered. If we want our theory to be fully self-consistent without reordering and capable of predicting any correlations to $\mathcal O(\chi^2)$, we should use the full equations \eqref{eqn:signalFULL} and \eqref{eqn:pumpFULL}. 
\end{document}